\documentclass[reprint,aps,prb]{revtex4-2}
\usepackage{amsmath}
\usepackage{amssymb}
\usepackage{braket}
\usepackage{graphicx}
\usepackage{hyperref}

\begin{document}

\title{Thermofield theory for finite-temperature electronic structure}

\author{Gaurav Harsha}
% \email{gharsha@umich.edu}
\affiliation{Department of Chemistry, University of Michigan, Ann Arbor, Michigan 48105}
\author{Thomas M. Henderson}
\affiliation{Department of Physics and Astronomy, Rice University, Houston, Texas 77005}
\affiliation{Department of Chemistry, Rice University, Houston, Texas 77005}
\author{Gustavo E. Scuseria}
\affiliation{Department of Physics and Astronomy, Rice University, Houston, Texas 77005}
\affiliation{Department of Chemistry, Rice University, Houston, Texas 77005}

\begin{abstract}
	Wave-function methods have offered a robust, systematically improvable means to study ground-state properties in quantum many-body systems. Theories like coupled cluster and their derivatives provide highly accurate approximations to the energy landscape at a reasonable computational cost. Analogs of such methods to study thermal properties, though highly desirable, have been lacking because evaluating thermal properties involve a trace over the entire Hilbert space, which is a formidable task. Besides, excited-state theories are generally not as well studied as ground-state ones.
	In this mini-review, we present an overview of a finite-temperature wave function formalism based on thermofield dynamics to overcome these difficulties. Thermofield dynamics allows us to map the equilibrium thermal density matrix to a pure state, i.e., a single wave function, albeit in an expanded Hilbert space. Ensemble averages become expectation values over this so-called thermal state.
	Around this thermal state, we have developed a procedure to generalize ground-state wave function theories to finite temperatures.
	As explicit examples, we highlight formulations of mean-field, configuration interaction, and coupled cluster theories for thermal properties of fermions in the grand-canonical ensemble.
	To assess the quality of these approximations, we also show benchmark studies for the one-dimensional Hubbard model, while comparing against exact results.
	We will see that the thermal methods perform similarly to their ground-state counterparts, while merely adding a pre-factor to the asymptotic computational cost.
	They also inherit all the properties, good or bad, from the ground-state methods, signifying the robustness of our formalism and the scope for future development.
\end{abstract}

\maketitle

\section{\label{sec1}Introduction}
The design and study of new materials and molecules is becoming heavily reliant on the ability to describe and predict their macroscopic properties from a quantum mechanical treatment of the underlying electronic structure.
Many conventional applications involve molecules and materials with optical gaps much larger than the temperature of the environment, e.g., large gap molecules and insulating solids at room temperatures.
The properties in such systems are governed solely by the ground state because the thermal energy is incapable of exciting enough electrons into higher energy states to make any practically observable difference.
On the other hand, for quantum systems with optical gaps comparable to, or smaller than, the thermal energy, the ground sate is no longer a sufficient description.
Accounting for thermal effects become essential for an accurate understanding of the electronic structure under these conditions, which often arise in novel applications based on strongly correlated materials (e.g., high-Tc superconductors and transition metal complexes at room temperature).~\cite{lee_doping_2006,raveau_strongly_2011}
Other examples include, but are not limited to, chemical reactions driven by hot-electrons~\cite{mukherjee_hot_2013} and those or in extreme geological environments.~\cite{guillot_interiors_1999-1,wann_high-temperature_2017}

Finite-temperature properties are defined as an average of all expected outcomes of an observable, weighted over an appropriate ensemble. For example, in the grand-canonical ensemble, the expectation value of an observable $A$ at inverse temperature $\beta=1/k_BT$ and chemical potential $\mu$ is defined as
\begin{equation}
	\braket{A} (\beta, \mu) = \frac{1}{\mathcal{Z}} \sum_{m} \braket{m | A | m} e^{-\beta (E_m - \mu N_m)},
	\label{eq:enesmble-avg-trace}
\end{equation}
where $\bra{m}$ and $\ket{m}$ are the Fock-space bra- and ket-eigenvectors of the system Hamiltonian $H$ with energy $E_m$ and particle number $N_m$, and $\mathcal{Z} = \sum_m e^{-\beta (E_m - \mu N_m)}$ is the partition function.
This expectation value can also be expressed as a trace of $A$ over the ensemble density matrix
\begin{equation}
	\braket{A} (\beta, \mu) = \mathrm{Tr} (\rho A),
\end{equation}
where $\rho = e^{-\beta (H-\mu N)} / \mathcal{Z}$ is the Gibbs-Boltzmann density operator, with $H$ and $N$ being the Hamiltonian and the total particle-number operator.
A direct application of conventional, state-specific, wave function methods to calculate thermal ensemble averages would be highly impractical because:
\begin{enumerate}
	\item The number of states in the Hilbert space that need to be counted in the ensemble grows exponentially with the system size.
	\item Approximating individual eigenstates of the Hamiltonian is a challenging task. While several reliable methods are now available for the ground-state, few excited state methods are equally accurate and efficient.
\end{enumerate}
As a result, the many-body problem becomes far more difficult at finite temperatures. Therefore, beyond mean-field methods,~\cite{mermin_stability_1963,sokoloff_consequences_1967} Matsubara perturbation theory~\cite{matsubara_new_1955} (generally employed in Green's function methods) as well as quantum Monte Carlo have been considered an important ingredient in a successful finite-temperature electronic structure method.

On the other hand, state-specific wave function theories offer more control, are systematically improvable, and often provide accurate results at a computational cost that grows only as a polynomial in system size.
For instance, Hartree-Fock (HF) theory is a mean-field method that generally provides a good qualitative result at $\mathcal{O}(n^3)$ computational scaling ($n$ being a measure of the system size). Configuration interaction (CI) with singles and doubles (CISD) and second-order M\o ller-Plesset perturbation theory (MP2) improve upon HF but require increased computational effort ($\mathcal{O}(n^6)$ for CISD and $\mathcal{O}(n^5)$ for MP2).
Coupled cluster (CC) theory,~\cite{crawford_introduction_2000,bartlett_coupled-cluster_2007} truncated at singles and doubles as well (CCSD), generally outperforms HF, CISD, and MP2 and scales as $\mathcal{O} (n^6)$.
Wave function methods have accumulated decades of experience, and several existing codes and numerical techniques are available to make them even more efficient.~\cite{sun_pyscf_2018,giannozzi_quantum_2020}
Because of such important qualities, the development of new wave function methods to study quantum mechanical systems at finite temperatures is highly desirable.
Consequently, finite-temperature wave function methods have seen a growing interest in recent years. Beyond the thermal mean-field theory,
several new methods have been proposed, e.g. Ancilla density matrix renormalization group (DMRG),~\cite{lichtenstein_finite-temperature_2001,verstraete_matrix_2004,feiguin_finite-temperature_2005,czarnik_variational_2016} minimally entangled typical thermal states,~\cite{white_minimally_2009,stoudenmire_minimally_2010} thermal density functional theory,~\cite{pittalis_exact_2011} perturbation theory methods,~\cite{santra_finite-temperature_2017,hirata_converging_2019} and CI- or CC-like techniques,~\cite{sanyal_thermal_1992,sanyal_thermal_1992,sanyal_systematic_1993,mandal_finite-temperature_2003,hermes_finite-temperature_2015,white_time-dependent_2018,hummel_finite_2018,harsha_thermofield_2019,harsha_thermal_cc_2019,shushkov_real-time_2019,harsha_wave_2020,harsha_thermal_2022} to name a few.

In this conference article, we present a short review of thermal wave function methods, with particular emphasis on a new formalism for generalizing existing ground-state wave function theories to finite-temperatures by using the theory of thermofield dynamics.~\cite{matsumoto_thermo_1983,semenoff_functional_1983,umezawa_methods_1984}
Most of the theory, as well as some key results, discussed here were reported by the authors in Refs.~\onlinecite{harsha_thermofield_2019,harsha_thermal_cc_2019,harsha_wave_2020,harsha_thermal_2022}.

\section{\label{sec2}Wave function methods at finite temperature}
While finite-temperature (or imaginary-time) generalizations of the time-dependent perturbation theory have been around for over half a century, one of the earliest ideas to generalize a wave function theory (coupled cluster, in this context) to finite-temperatures was proposed by Sanyal, Mukherjee and co-authors.
In a series of papers,~\cite{sanyal_thermal_1992,sanyal_thermal_1992,sanyal_systematic_1993,mandal_finite-temperature_2003} they developed the thermal cluster cumulant (TCC) theory, in which the interaction picture Gibbs-Boltzmann operator is parametrized using a CC-like exponential operator.
By using a thermal generalization of the Wick's theorem and associated normal ordering for this exponential ansatz, they showed that the free energy can be expressed using fully contracted term in the normal ordered expansion.

Recently, different ways to generalize the coupled cluster theory to finite temperatures have been sought separately by Hummel,~\cite{hummel_finite_2018} and White et al.~\cite{white_time-dependent_2018}
Both of these works rely on re-interpreting the coupled cluster diagrams from the perspective of time-dependent perturbation theory on the imaginary-time axis.
While Hummel and White's approaches differ in inspiration, by using the time-dependent perturbation theory, they end up relying on the thermal Wick's theorem, much like in Sanyal and Mukherjee's TCC theory.
Consequently, up to difference in implementation, Hummel, Sanyal, and White's finite-temperature coupled cluster result in similar working equations.

Another formalism for thermal wave function methods, inspired by thermofield dynamics, was proposed by the authors of this review.
The thermofield approach relies on an explicit wave function representation of the Gibbs-Boltzmann density operator, which allows a direct generalization of any ground-state wave function ansatzes to finite temperatures.

Using thermofield dynamics or related principles to construct thermal wave function theories is not new.
A connection of the TCC with the thermofield dynamics was explored by Sanyal and Mukherjee in Refs.~\citenum{sanyal_thermal_1992,sanyal_systematic_1993}.
They later noted that the thermofield theory was not a necessary ingredient in TCC.
Ancilla DMRG also builds on techniques similar to thermofield dynamics for purification of the ensemble density matrix.
Furthermore, a coupled cluster formalism for finite-temperature dynamics, inspired by thermofield dynamics, was also explored by Shushkov et al. in Ref.~\citenum{shushkov_real-time_2019}.
However, to our knowledge, a general framework was proposed only in Ref.~\citenum{harsha_thermofield_2019}.

\begin{figure*}[tb]
	\centering
	\includegraphics[width=0.7\textwidth]{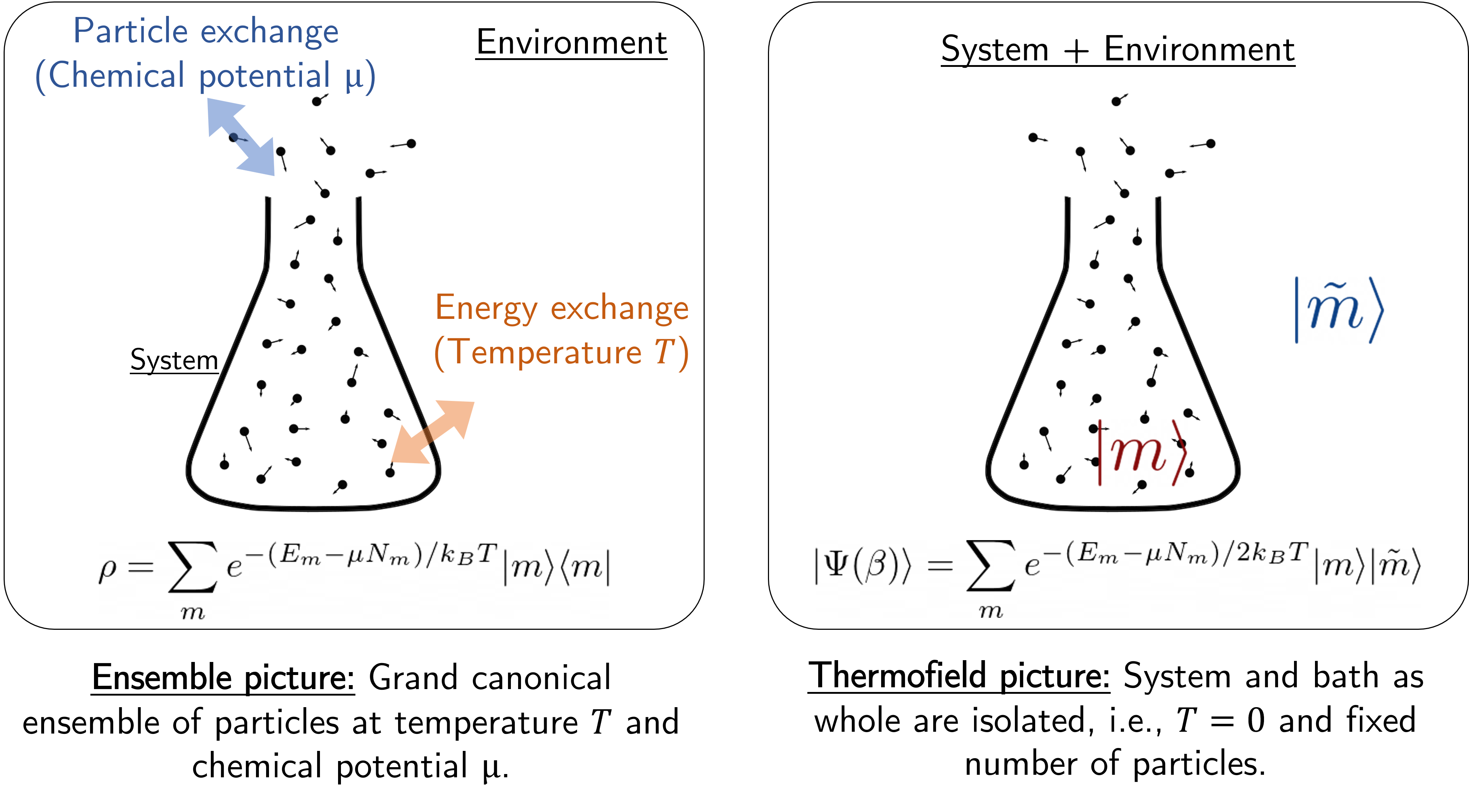}
	\caption{Physical interpretation of how a single wave function in an enlarged space in the thermofield framework can encode the thermal ensemble density matrix}
	\label{fig:thermofield}
\end{figure*}

\section{\label{sec3}Thermofield Dynamics}
Thermofield dynamics was originally proposed to study time-dependent phenomena at finite-temperature using path integrals, much like the Keldysh approach~\cite{keldysh_diagram_1965} in the Matsubara imaginary time formalism.
It provides a way to construct a pure-state representation for the thermal density matrix. That is, we can write down a \emph{thermal wave function} $\ket{\Psi (\beta, \mu)}$, such that the trace over an ensemble of states in Eq. \ref{eq:enesmble-avg-trace} can be expressed as an expectation value over the thermal state, i.e.,
\begin{equation}
  \braket{A} = \frac{
		\braket{\Psi (\beta, \mu) | A | \Psi (\beta, \mu)}
	}{
		\langle \Psi (\beta, \mu) \vert \Psi (\beta, \mu) \rangle
	}.
  \label{eq:ensemble-avg-thermofield}
\end{equation}
Such a purification of the ensemble density matrix is achieved by working in an enlarged space that is made up from
\begin{itemize}
	\item the original Hilbert space $\mathcal{H}$,
	\item and its complex conjugate copy $\tilde{\mathcal{H}}$, also known as the \emph{tilde-conjugate} space.
\end{itemize}
The thermal state is defined as
\begin{equation}
	\ket{\Psi (\beta, \mu)} = \frac{1}{\mathcal{Z}} \sum_m e^{-\beta(E_m - \mu N_m) / 2} \ket{m} \ket{\tilde{m}},
\end{equation}
where, once again, $\beta = 1 / k_B T$ is the inverse temperature, $\mathcal{Z}$ is the partition function, $\ket{m}$ is an eigenstate of the Hamiltonian $H$ with eigenvalue $E_m$, and $N_m$ number of particles, and $\ket{\tilde{m}}$ is the conjugate-space counterpart of $\ket{m}$.
In the grand-canonical ensemble, the thermal state can be explicitly written as
\begin{subequations}
  \label{eq:thermal-state}
	\begin{align}
  	\ket{\Psi (\beta, \mu)} &= e^{-\beta(H - \mu N)/2} \ket{\mathbb{I}},
		\\
		\ket{\mathbb{I}} &= \prod_p \left(1 + c_p^\dagger \tilde{c}_p^\dagger \right) \ket{-;-},
	\end{align}
\end{subequations}
where $p$ is a spin-orbital index, $c_p^\dagger$ creates a particle in the $p$th physical spin-orbital and $\tilde{c}^\dagger_p$ creates a particle in the corresponding conjugate spin-orbital. Finally, $\ket{-; -}$ denotes the vacuum for physical and conjugate spaces.
The state $\ket{\mathbb{I}}$ is also the state with maximal entanglement between $\mathcal{H}$ and $\tilde{\mathcal{H}}$ and is the exact thermal state at infinite temperature (i.e. $\beta=0$) and $\mu=0$.

A simple physical interpretation for the thermofield approach is that by introducing the \emph{tilde-conjugate} degrees of freedom, the system and the thermal bath (or the environment) are treated equally and as one big, isolated system. The thermal state can be thought of as the wave function for this combined macro system. Since the dimensionality of a typical quantum system is much smaller than that of the environment, we can use the Schmidt decomposition~\cite{schmidt_zur_1907} to factorize the thermal state in the form of Eq. \ref{eq:thermal-state}.
Figure \ref{fig:thermofield} provides a graphical depiction of this interpretation.
Thermofield dynamics has already shown great potential in the study many-electron systems in chemical and condensed-matter systems.\cite{suzuki_thermo_1985,hatsuda_mean_1989,walet_thermal_1990,de_vega_thermofield-based_2015,borrelli_quantum_2016,nocera_symmetry-conserving_2016,chen_finite_2017,borrelli_simulation_2017,wu_variational_2019}

\section{\label{sec4}Finite-temperature wave function methods}
By definition, the grand-canonical thermal state in Eq. \ref{eq:thermal-state} satisfies the following evolution equations in the inverse temperature $\beta$ and chemical potential $\mu$:
\begin{subequations}
	\label{eq:evolution}
	\begin{align}
		\frac{\partial \ket{\Psi (\beta, \mu)}}{\partial \beta}
		&= -\frac{1}{2} H \ket{\Psi (\beta, \mu)},
		\label{eq:beta-evolution}
		\\
		\frac{\partial \ket{\Psi (\beta, \mu)}}{\partial \mu}
		&= \frac{\beta}{2} N \ket{\Psi (\beta, \mu)}.
		\label{eq:mu-evolution}
	\end{align}
\end{subequations}
An exact solution of these evolution equations is equivalent to exact diagonalization (or Full CI) in the Fock space.
However, since we only need to work with a single wave function, we can use any conventional ground-state wave function approximation to solve and integrate the imaginary time ($\beta$) and chemical potential ($\mu$) evolution equations.
Here, we will show explicit constructions for the mean-field, CI and CC approximations.

\subsection{Mean-field theory}
In the mean-field approximation, we use an effective one-body Hamiltonian $H_0$ to integrate the evolution equations in Eq. \ref{eq:evolution}. As long as $H_0$ does not depend on $\beta$ and $\mu$, the integration can be performed analytically and the mean-field thermal state $\ket{\Phi (\beta, \mu)}$ is given by
\begin{equation}
	\ket{\Phi (\beta, \mu)} = e^{-\beta(H_0 - \mu N) / 2} \ket{\mathbb{I}}.
	\label{eq:mean-field-thermal-state}
\end{equation}
For electronic systems, a convenient choice for $H_0$ is the sum of Fock operators from the ground-state Hartree-Fock state. In the Hartree-Fock molecular orbital basis, in which the Fock matrix is diagonal and $H_0 = \sum_p \epsilon_p c_p^\dagger c_p$, the normalized mean-field thermal state can be factorized as
\begin{equation}
	\ket{\Phi (\beta, \mu)} = \prod_{p} \left(
		u_p + v_p c_p^\dagger \tilde{c}_p^\dagger
	\right) \ket{-;-},
	\label{eq:fock-mean-field-thermal-state}
\end{equation}
where $u_p = 1 / \sqrt{1 + e^{-\beta (\epsilon_p - \mu)}}$, and $v_p = \sqrt{1 - u_p^2}$. Here, the product runs over all spin-orbitals $p$.
In order to maintain charge neutrality, the chemical potential $\mu$ is adjusted for every $\beta$ grid-point.
The mean-field thermal state in Eq.~\ref{eq:fock-mean-field-thermal-state} is a Bardeen-Cooper-Schrieffer (BCS) state which reduces to the ground-state Hartree-Fock in the zero-temperature limit where $v_\text{occupied} = 1 = u_\text{virtual}$, and $v_\text{virtual} = 0 = u_\text{occupied}$.
The thermal analogue of Cooper pairs are formed between the physical and auxiliary orbital pairs $\{p, \tilde{p}\}$.

\subsection{Correlated methods}
For an accurate description of quantum mechanical systems, we need to go beyond mean-field theory and capture the effects of correlation in the system.
Traditional ground-state methods such as perturbation theory, CI, and CC are some of the most popular and effective methods for this task.
In general, a correlated approximation to the thermal state is constructed as
\begin{equation}
	\ket{\Psi (\beta, \mu)} = \Gamma (\beta, \mu) \ket{\Phi (\beta, \mu)},
	\label{eq:wave-operator}
\end{equation}
where the wave operator $\Gamma (\beta, \mu)$ adds correlation atop the mean-field reference.
The wave operator $\Gamma (\beta, \mu)$ is constructed using quasiparticle excitation operators for the mean-field reference $\ket{\Phi (\beta, \mu)}$, in the same way as particle-hole excitations for a Hartree-Fock reference are used in ground-state theories.
These thermal quasiparticles operators, $\{a_p, a_p^\dagger, \tilde{a}_p, \tilde{a}_p^\dagger\}$ are defined in such a way that
\begin{equation}
	a_p (\beta, \mu) \ket{\Phi (\beta, \mu)} = 0 = \tilde{a}_p (\beta, \mu) \ket{\Phi (\beta, \mu)}.
\end{equation}
For the grand canonical mean-field state, defined in Eq. \ref{eq:fock-mean-field-thermal-state}, the fermion and quasiparticle creation and annihilation operators are related via a Bogoliubov transformation:
\begin{equation}
  \begin{bmatrix}
    a_p  \\
    \tilde{a}_p^\dagger
  \end{bmatrix} =
  \begin{bmatrix}
    u_p & -v_p\\
    v_p & u_p
  \end{bmatrix}
  \begin{bmatrix}
    c_p\\
    \tilde{c}_p^\dagger
  \end{bmatrix}.
  \label{eq:bogoliubov}
\end{equation}
The $\beta$- and $\mu$-evolution equations for the correlated wave function becomes
\begin{subequations}
	\begin{align}
		\frac{\partial \Gamma}{\partial \beta} \ket{\Phi}
		&=
		-\frac{1}{2} (H \Gamma - \Gamma H_0) \ket{\Phi},
		\\
		\frac{\partial \Gamma}{\partial \mu} \ket{\Phi}
		&=
		\frac{\beta}{2} (N \Gamma - \Gamma N) \ket{\Phi},
	\end{align}
\end{subequations}
where both $\Gamma$ and $\ket{\Phi}$ depend on $\beta$ and $\mu$, as noted in Eq.~\ref{eq:wave-operator}.
Comparing different orders of quasiparticle excitations on the left and right hand side of this equation defines our initial value problem. Note that at $\beta = 0$, mean-field theory is exact, and $\Gamma (\beta = 0, \mu) = 1$.

\subsubsection{Configuration interaction theory}
For the CI approximation, the wave operator $\Gamma (\beta, \mu)$ is expressed as a linear combination of various quasiparticle excitations.
At the level of single and double excitations (CISD), the thermal state is defined as
\begin{equation}
	\ket{\Psi_{CI} (\beta, \mu)}
	=
	e^{t_0} (1 + T) \ket{\Phi (\beta, \mu)},
	\label{eq:ci-state}
\end{equation}
where $t_0$ is a scalar that keeps track of the norm, and the excitation operator $T$ is defined as
\begin{align}
	T &=
	\sum_{pq} t_{pq} a_p^\dagger (\beta, \mu) \tilde{a}_q^\dagger (\beta, \mu)
	\nonumber
	\\
	& + \frac{1}{4} \sum_{pqrs} t_{pqrs} a_p^\dagger (\beta, \mu) a_q^\dagger (\beta, \mu) \tilde{a}_s^\dagger (\beta, \mu) \tilde{a}_r^\dagger (\beta, \mu).
	\label{eq:ci-operator}
\end{align}
While $t_0$, $t_{pq}$, and $t_{pqrs}$ also depend on $\beta$ and $\mu$, we have suppressed the explicit dependence for brevity.

\subsubsection{Coupled cluster theory}
In coupled cluster theory, the wave operator is expressed as an exponential of the quasiparticle excitations. At the level of singles and doubles approximation (CCSD), the thermal state is defined as
\begin{equation}
	\ket{\Psi_{CC} (\beta, \mu)}
	=
	e^{t_0 + T} \ket{\Phi (\beta, \mu)},
	\label{eq:cc-state}
\end{equation}
where $t_0$ is a scalar and $T$ has the same form as in Eq. \ref{eq:ci-operator}. Technical details on the derivation and exact form of evolution equations for the amplitudes $t_{pq}$ and $t_{pqrs}$ in CISD and CCSD can be found in Refs.~\citenum{harsha_thermofield_2019,harsha_thermal_cc_2019}.

\begin{figure*}[tb]
	\centering
	\includegraphics[width=0.45\textwidth]{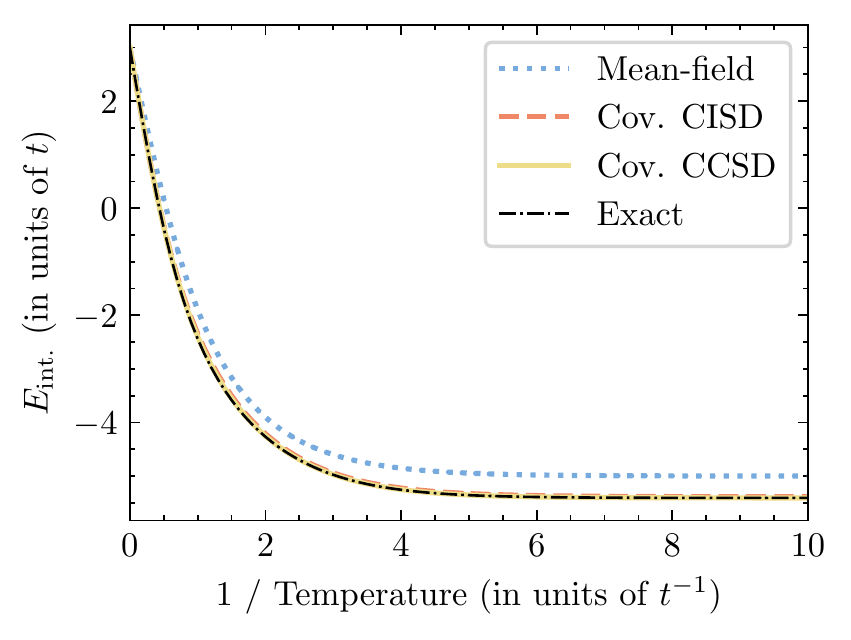}
	\includegraphics[width=0.46\textwidth]{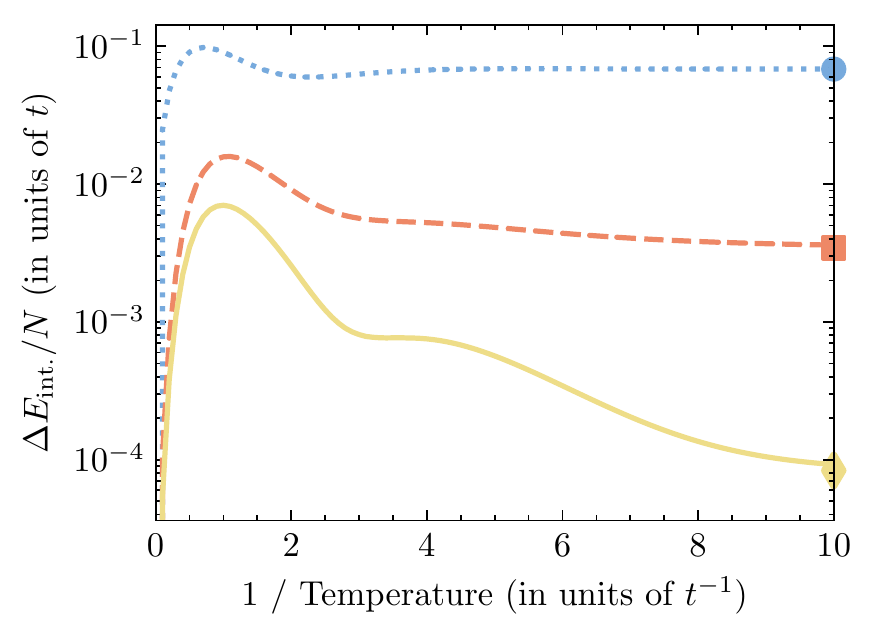}
	\caption{
		\textbf{Internal energy and associated error}: Comparison of (left) internal energy for approximate and exact thermal wave functions, and (right) corresponding errors in the approximate theories (mean-field, CISD, and CCSD) with respect to exact internal energy for the 6-site Hubbard model with $U/t = 2$ at half-filling on average. The markers on the y-axis in the error plot show the errors in the ground-state mean-field, CISD and CCSD.
		The results are similar to Fig. 1b of Ref.~\citenum{harsha_thermal_cc_2019}, but have been re-computed for this review.
	}
	\label{fig:hubbard-6site-u2}
\end{figure*}

\subsubsection{Computational scaling and accuracy}
The asymptotic computational scaling of thermal wave function methods is the same as their ground-state counterparts, albeit with a pre-factor proportional to the grids used in integration along $\beta$- and $\mu$-directions.
Therefore, while finite-temperature mean-field scales as $\mathcal{O}(n^3)$, correlated methods such as CISD and CCSD scale as $\mathcal{O} (n^6)$, where $n$ is the number of spin-orbitals used to describe the system.
Additionally, the scaling of ground-state methods often benefit from a clear distinction between the occupied and unoccupied (or virtual) orbitals, e.g., ground-state CCSD scales as $\mathcal{O}(n_o^2 n_v^4)$, where $n_o$ and $n_v$ are the number of occupied and virtual orbitals.
At finite temperatures, all orbitals are treated equally, and a distinction based on orbital occupation does not exist.
In fact, considering that finite-temperature mean-field is a BCS-style state, thermal CI and CC are equivalent in formulation and scaling to ground-state quasiparticle CI and CC.~\cite{henderson_quasiparticle_2014}

While higher order approximations have not been implemented so far, we expect that the thermal CI and CC will become more accurate as quasiparticle excitations at the level of triples, quadruples, etc. are incorporated.
However, it is interesting to note that in the grand-canonical ensemble, finite-temperature CI and CC will become exact only if quasiparticle excitations up to the order equal to the number of spin-orbitals are used.
For instance, in Ref.~\citenum{harsha_thermal_cc_2019}, we showed that while ground-state CCSD and CISD are exact for two-electron systems, finite-temperature CISD and CISD are not.
This again is a property inherited from quasiparticle CI and CC.~\cite{henderson_quasiparticle_2014}
 
\subsection{Physical properties}
With the correlated thermal states, the finite-temperature expectation value of a physical observable, represented by an operator $A$, can be computed using Eq.~\ref{eq:ensemble-avg-thermofield}. For instance, the CI approximation to $\braket{A} (\beta, \mu)$ is defined as
\begin{equation}
	\braket{A}_{CI} (\beta, \mu) = \frac{
		\braket{\Psi_{CI} (\beta, \mu) | A | \Psi_{CI} (\beta, \mu)}
	}{
		\braket{\Psi_{CI} (\beta, \mu) | \Psi_{CI} (\beta, \mu)}
	}.
	\label{eq:ci-expectation-value}
\end{equation}
On the other hand, for the coupled cluster method, a symmetric expectation value is generally not feasible. Taking inspiration from ground-state CC, we use a linear, or CI, approximation for the \textit{bra} thermal state, i.e.,
\begin{subequations}
	\begin{align}
		\bra{\Psi_{CI} (\beta, \mu)}
		&= \bra{\Phi (\beta, \mu)} \left(1 + W (\beta, \mu)\right) e^{w_0},
		\\
		&= \bra{\Phi (\beta, \mu)} \left(1 + Z(\beta, \mu)\right) e^{z_0 - T(\beta, \mu)},
	\end{align}
\end{subequations}
where the operators $W (\beta, \mu)$ and $Z(\beta, \mu)$ are quasiparticle de-excitation operators (excitations on the \textit{bra}), and are defined as
\begin{subequations}
	\begin{align}
		W(\beta, \mu)
		&= \sum_{pq} w_{pq} \tilde{a}_q (\beta, \mu) a_p (\beta, \mu)
		\nonumber
		\\
		+ \frac{1}{4} \sum_{pqrs} & w_{pqrs} \tilde{a}_r (\beta, \mu) \tilde{a}_s (\beta, \mu) a_q (\beta, \mu) a_p (\beta, \mu),
		\\
		Z(\beta, \mu)
		&= \sum_{pq} z_{pq} \tilde{a}_q (\beta, \mu) a_p (\beta, \mu)
		\nonumber
		\\
		+ \frac{1}{4} \sum_{pqrs} & z_{pqrs} \tilde{a}_r (\beta, \mu) \tilde{a}_s (\beta, \mu) a_q (\beta, \mu) a_p (\beta, \mu).
	\end{align}
\end{subequations}
Once again, the explicit $\beta$- and $\mu$-dependence in the $w$- and $z$-amplitudes has been suppressed for brevity.
Using the fact that $\bra{\Phi (\beta, \mu)} e^{T (\beta, \mu)} = \bra{\Phi (\beta, \mu)}$, a transformation between $Z$ and $W$ can be derived:
\begin{widetext}
	\begin{subequations}
		\begin{align}
			\bra{\Phi (\beta, \mu)} (1 + W (\beta, \mu)) e^{w_0}
			&= \bra{\Phi (\beta, \mu)} e^{-T(\beta, \mu)} (1 + W (\beta, \mu)) e^{w_0} e^{T(\beta, \mu)} e^{-T (\beta, \mu)},
			\\
			\Rightarrow \bra{\Phi (\beta, \mu)} (1 + Z (\beta, \mu)) e^{z_0}
			&= \bra{\Phi (\beta, \mu)} e^{-T(\beta, \mu)} (1 + W (\beta, \mu)) e^{w_0} e^{T(\beta, \mu)}.
		\end{align}
	\end{subequations}
	Finally, with a linearized approximation to the \textit{bra} thermal state, we define the finite-temperature CC expectation values as
	\begin{subequations}
		\label{eq:cc-expectation-value}
		\begin{align}
			\braket{A}_{CC} (\beta, \mu)
			&= \frac{
				\braket{\Psi_{CI} (\beta, \mu) | A | \Psi_{CC} (\beta, \mu)}
			}{
				\braket{\Psi_{CI} (\beta, \mu) | \Psi_{CC} (\beta, \mu)}
			},
			\\
			&= \frac{
				\braket{
					\Phi (\beta, \mu) |
					\left( 1 + Z(\beta, \mu) \right) e^{-T(\beta, \mu)} A e^{T(\beta, \mu)}
					| \Phi (\beta, \mu)
				}
			}{
				\braket{\Phi (\beta, \mu) | \Phi (\beta, \mu)}
			}.
		\end{align}
	\end{subequations}
	The grand potential cannot be computed as an operator expectation value. However, it is directly related to the partition function which, in thermofield dynamics, is given by the normalization of the thermal wave function. The finite-temperature coupled cluster grand potential can thus be computed as
	\begin{subequations}
		\label{eq:grand-pot}
		\begin{align}
			\Omega_{CC} &= \Omega_0 + \Omega_c,
			\\
			\Omega_0 &= -\frac{1}{\beta} \log
			\braket{\mathbb{I} | e^{-\beta (H_0 - \mu N)} | \mathbb{I}}
			= -\frac{1}{\beta} \log \left[
				\prod_p \left(1 + e^{-\beta (\epsilon_p - \mu)}\right)
			\right]
			,
			\label{eq:mean-field-grand-pot}
			\\
			\Omega_c &= -\frac{1}{\beta} \log 
			\left[
				\frac{
					\braket{\Phi (\beta, \mu) | (1 + Z (\beta, \mu)) e^{z_0} e^{-T(\beta, \mu)} e^{t_0 + T(\beta, \mu)} | \Phi (\beta, \mu)}
				}{
					\braket{\Phi (\beta, \mu) | \Phi (\beta, \mu)}
				}
			\right]
			= -\frac{1}{\beta} \left(t_0 + z_0\right).
		\end{align}
	\end{subequations}
\end{widetext}
Here, $\Omega_0$ and $\Omega_c$ denote the the mean-field and correlation contributions to the grand potential. In simplifying Eq.~\ref{eq:mean-field-grand-pot}, we recall our definition for the mean-field Hamiltonian, $H_0 = \sum_p \epsilon_p c_p^\dagger c_p$.
Other thermodynamic properties can be derived from the grand potential. For example, given the coupled cluster internal energy $E = \braket{H}_{CC}$ and particle number $N = \braket{N}_{CC}$, the entropy is defined as:
\begin{equation}
	S_{CC} = \frac{1}{T} \left(
		\braket{H}_{CC} - \mu \braket{N}_{CC} - \Omega_{CC}
	\right),
	\label{eq:entropy-expectation}
\end{equation}
where the internal energy, $E = \braket{H}_{CC}$, and the number of particles, $N = \braket{N}_{CC}$, are simply the thermal expectation values of the Hamiltonian and the number operator,
We can also define the particle number, entropy, and the internal energy as derivatives of the grand potential:
\begin{subequations}
	\label{eq:grand-pot-derivatives}
	\begin{align}
		N &= -\frac{\partial \Omega}{\partial \mu},
		\\
		S &= \beta^2 \frac{\partial \Omega}{\partial \beta},
		\\
		E &= \Omega + \mu N + \frac{1}{\beta} S.
	\end{align}
\end{subequations}
In principle, the derivatives of the grand potential (Eq.~\ref{eq:grand-pot-derivatives}) and the expectation values (e.g., Eq.~\ref{eq:entropy-expectation}) will result in properties that are numerically identical only when the thermal \textit{bra} and \textit{ket} make the grand potential stationary.~\cite{harsha_thermal_2022}
For our finite-temperature coupled cluster theory, where we time-evolve the CI-like \textit{bra} and the CC-like \textit{ket} separately, the stationarity of the grand potential is not guaranteed.
However, in practice, the derivative and expectation value formalisms lead to very similar results (see, for example, the comparison of magnetization properties in the one-dimensional transverse field Ising model in Ref.~\citenum{harsha_thermal_2022})

\section{\label{sec5}Results}
To study the accuracy of various approximations to the thermal state, we use the one-dimensional Hubbard model~\cite{hubbard_electron_1963} as a benchmark system, the Hamiltonian for which is defined as
\begin{equation}
  H = -t \sum_{\langle p,q \rangle,  \sigma} \left(c_{p,\sigma}^\dagger \, c_{q,\sigma} + \textrm{h.c.}\right) + U \, \sum_{p} n_{p,\uparrow} \, n_{p,\downarrow},
	\label{eq:hubbard}
\end{equation}
where $\langle,\rangle$ denotes that the sum is carried over connected lattice sites, $t$ denotes the hopping strength, $U$ denotes the strength of the on-site Coulomb repulsion, and $n_{p,\sigma} = c_{p,\sigma}^\dagger \, c_{p,\sigma}$ is the number operator for lattice site $p$ and spin $\sigma$.
The physics of the Hubbard model depends only on the ratio $U/t$, which characterizes the correlation strength.

The left panel in Fig. \ref{fig:hubbard-6site-u2} shows the total internal energy for the mean-field theory, CISD, CCSD, and exact thermal wave functions for the 6-site Hubbard model with $U/t = 2$ at half-filling on average. The right panel shows the errors in internal energy for the approximate methods against the exact result.
The error plot clearly show that thermal CISD and CCSD significantly improve the results as compared to the mean-field theory, with CCSD outperforming the other two.
The markers in the y-axis of the error plot indicate the errors in the corresponding ground-state theories, which establishes that the thermal wave function methods approach the ground-state methods in the limit $\beta \rightarrow \infty$ (or $T \rightarrow 0$).

\begin{figure}
	\centering
	\includegraphics[width=0.45\textwidth]{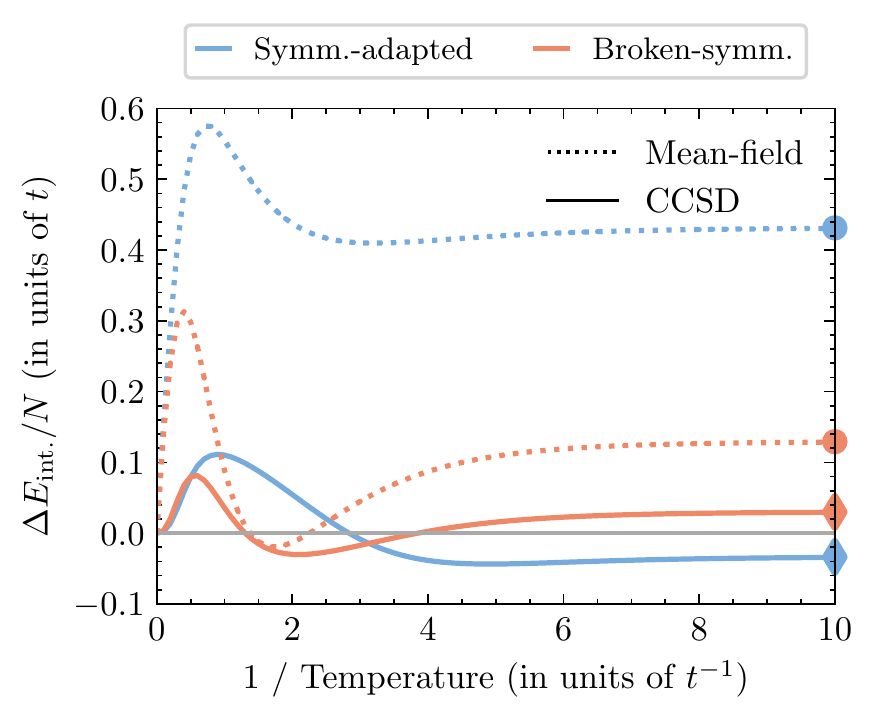}
	\caption{
		\textbf{Effect of mean-field reference}: Internal energy errors for thermal mean-field and CCSD based on symmetry-adapted and broken-symmetry Hartree-Fock eigenvalues to construct $H_0$. The system is 6-site Hubbard model with $U/t = 5$ and half-filling on average. The markers on the y-axis show the errors in the corresponding ground-state mean-field and CCSD methods.
		The results are similar to Fig. 5 of Ref.~\citenum{harsha_thermal_cc_2019}, but have been re-computed for this review.
	}
	\label{fig:rhf-vs-uhf}
\end{figure}

A common feature in ground-state methods is that the results in the correlated methods depend on the choice of mean-field reference state, e.g., a symmetry adapted restricted Hartree-Fock vs. a broken-symmetry unrestricted Hartree-Fock state. Figure \ref{fig:rhf-vs-uhf} shows a similar behavior for thermal wave function methods.
Here, we have used the 6-site Hubbard model with $U/t = 5$ at half-filling on average. The symmetry adapted thermal mean-field was constructed using zero-temperature restricted Hartree-Fock eigenvalues and orbitals as our $H_0$, while the broken symmetry mean-field was constructed from zero-temperature unrestricted Hartree-Fock.
The coupled cluster results were built upon the respective reference states.
The results once again show that thermal CCSD improves significantly over the respective mean-field reference, although the results depend on the choice of mean-field reference.
Both the symmetry adapted and broken-symmetry thermal mean-field and CCSD approach their ground-state counterparts as we approach zero temperature.

\begin{figure*}[tb]
	\centering
	\includegraphics[width=0.44\textwidth]{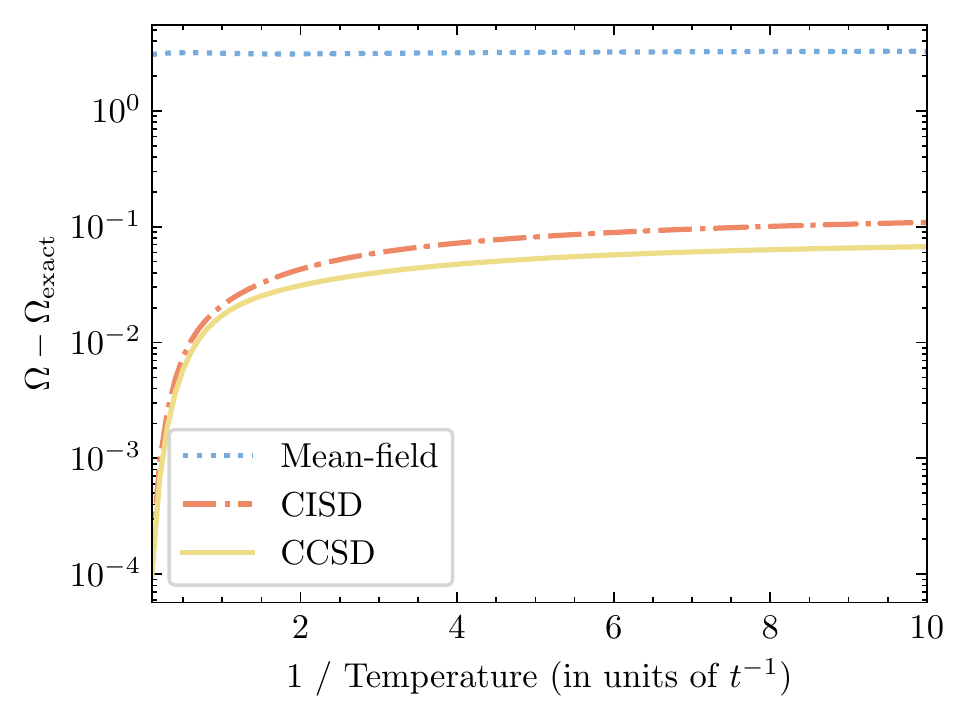}
	\includegraphics[width=0.45\textwidth]{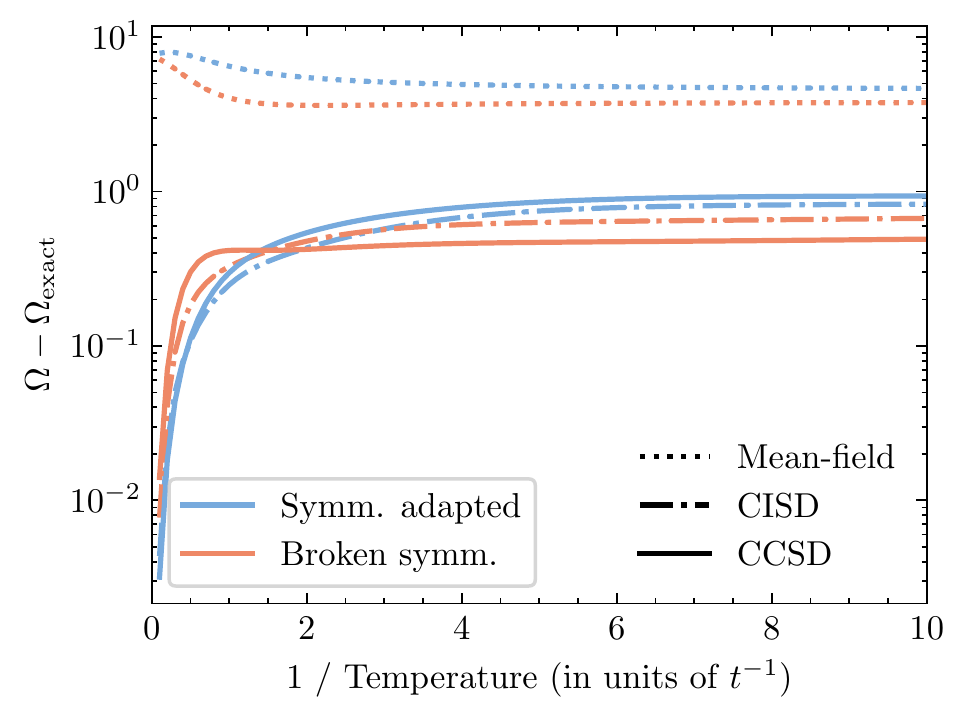}
	\caption{\textbf{Grand potential}: Error in grand potential for approximate thermal wave function theories (mean-field, CISD, and CCSD) with respect to exact values computed at chemical potential $\mu = 0$ for the 6-site Hubbard model with (left) $U/t = 2$, and (right) $U/t = 5$. For the latter, we show results for CISD and CCSD built upon a symmetry adapted and a broken symmetry thermal mean-field reference.}
	\label{fig:hubbard-grand-potential}
\end{figure*}

In Fig.~\ref{fig:hubbard-grand-potential}, we also compare mean-field, CISD, and CCSD approximations to the grand potential for 6-site Hubbard models at $U/t = 2$ and at $U/t=5$. For the latter, we use both the symmetry adapted and broken symmetry thermal mean-field reference, similar to Fig.~\ref{fig:rhf-vs-uhf}. For the comparison of grand potentials, we fix the chemical potential ($\mu = 0$) rather than fixing the total number of electrons.
As expected, we find that both CISD and CCSD improve significantly over the mean-field grand potential.
While other mean-field observables such as internal energy, entropy, etc. are, by definition, exact in the limit $\beta \rightarrow 0$ (i.e., infinite temperature), the grand potential $\Omega_0$ defined in Eq.~\ref{eq:mean-field-grand-pot} is not exact in the high-temperature limit.
This is because $\Omega_0$ includes contribution only from the mean-field Hamiltonian $H_0$.
For $U/t = 5$, it is well known that ground-state CCSD, whether symmetry adapted or symmetry broken, struggles to recover the correlation energy. This ineffectiveness of single-reference methods like CI and CC is also visible at finite temperatures.

\begin{figure}[b]
	\centering
	\includegraphics[width=0.45\textwidth]{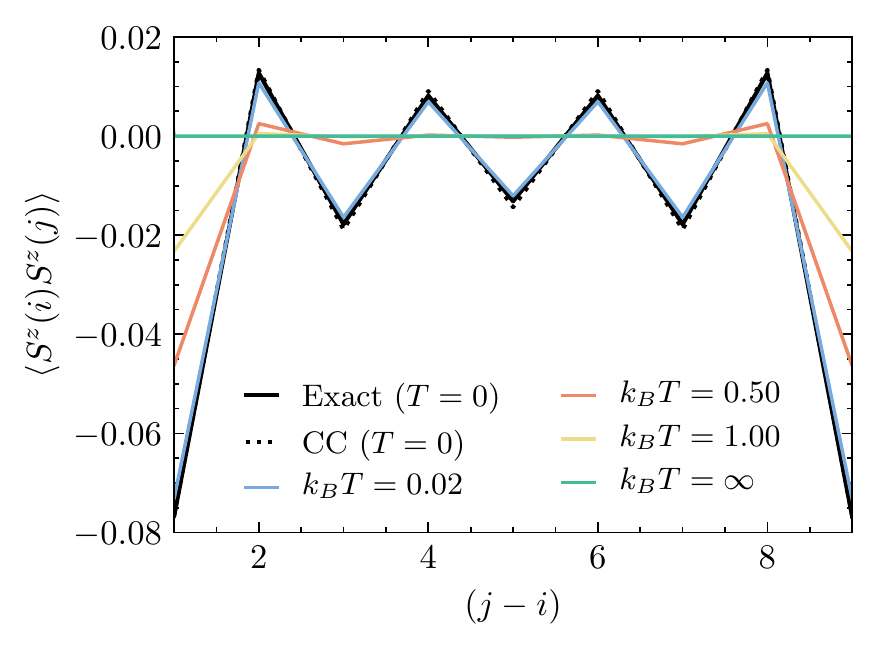}
	\caption{
		\textbf{Correlation function}: Temperature dependence of the spin-spin correlation function in the 10-site Hubbard model with $U/t = 2$.
		The results are similar to Fig. 6 of Ref.~\citenum{harsha_thermal_cc_2019}, but have been re-computed for this review.
	}
	\label{fig:sz-sz}
\end{figure}

At finite temperatures, we are more interested in properties other than internal energy. For example, in the one-dimensional Hubbard model, the spin-spin correlation function, defined as
\begin{equation}
  \chi(i,j) = \langle S^z(i)\:S^z(j) \rangle
\end{equation}
is an observable quantification of the correlation in the system. For a weakly interacting Hubbard model, the strength of this correlation function decays very rapidly with the distance between two lattice sites. On the other hand, as the interaction strength is increased, the length-scales over which $\chi(i,j)$ decays increases.
We use the biorthogonal framework outlined in Eq.~\ref{eq:cc-expectation-value} to evaluate these correlation functions.
Figure \ref{fig:sz-sz} shows the trends in $\chi(i, j)$ at various temperatures for the 10-site Hubbard model with $U/t = 2$. A fixed chemical potential was used for this calculation which ensures near-half-filling for all temperature points used.
The correlation function for exact and coupled cluster calculations for the ground-state (i.e., at $T = 0$) is also shown for reference.
We can observe that as the temperature is increased, the correlation strength as well as the correlation length decreases, which establishes that thermal CCSD is capable of accurately describing trends in the order parameters as a function of temperature.

\subsection{Zero-temperature limit of thermal methods}
Figures~\ref{fig:hubbard-6site-u2} and \ref{fig:rhf-vs-uhf} show that all the approximate finite-temperature methods converge to their ground-state counterparts in the zero-temperature limit.
However, the well behaved zero-temperature convergence of these methods arises from the fact that the thermal mean-field converges to the appropriate reference determinant, i.e.
\begin{equation}
	\lim_{\beta \rightarrow \infty} \ket{\Phi (\beta)} = \ket{\Phi_0},
\end{equation}
where $\ket{\Phi_0}$ is the ground-state reference Slater determinant.
This limit is well behaved only if the ground-state of the mean-field driver Hamiltonian $H_0 = \sum_p \epsilon_p c^\dagger_p c_p$ is non-degenerate. Otherwise, the zero-temperature limit of $\ket{\Phi (\beta)}$ is not guaranteed to be a single Slater determinant, and consequently, the correlated methods would not necessarily converge to single-reference ground-state counterparts.
For a repulsive Hamiltonian, even if the exact ground-state is degenerate, it is known that the general Hartree-Fock eigenvalues will not be degenerate.~\cite{bach_there_1994}
We would also like to note that degeneracies caused by spin are generally not a problem, because $H_0$ is defined by a zero-temperature Fock operator which tends to lift these degeneracies even for unrestricted or restricted open-shell Hartree-Fock.
Consider, for example, the hydrogen atom where we have placed the electron in the $1s \uparrow$ spin state, whose orbital energy is given by
\begin{equation}
	\epsilon_{1s \uparrow} = \braket{1s \uparrow | h | 1s \uparrow}
\end{equation}
since the self-Coulomb is cancelled by the self-exchange in Hartree-Fock theory.
On the other hand, the orbital energy of the $1s \downarrow$ state is raised by the Coulomb interaction with the $1s \uparrow$ orbital:
\begin{equation}
	\epsilon_{1s \downarrow} = \braket{1s \downarrow| h |1s \downarrow}
	+ \braket{1s \downarrow, 1s \uparrow | v | 1s \downarrow, 1s \uparrow}.
\end{equation}
Here, $h$ and $v$ denote the one- and two-electron interaction terms in the Hamiltonian.
Because the two orbitals are not degenerate, the zeroth-order Hamiltonian has a non-degenerate ground state to which, as $\beta \rightarrow \infty$, the thermal mean-field state converges.
In this case, our thermal mean-field will converge to the $\uparrow$-spin ground state.

Furthermore, in most scenarios with near-degeneracies in the mean-field Hamiltonian, ground-state single-reference coupled cluster breaks down, and the question of a ground-state limit for the finite-temperature method does not arise.
In such cases, we observe a divergence in the thermal CC amplitudes as we evolve towards zero temperature.

\section{\label{sec6}Connections with other finite-temperature CC theories}
In Section~\ref{sec2}, we introduced thermal CC theories by Sanyal et al., Hummel, and White et al., all of which are inspired by thermal Wick's theorem and imaginary-time-dependent perturbation theory.
While the thermofield version is completely different in its formulation, the resulting equations for the finite-temperature CC are identical to the other three methods, provided that other implementation details, such as the mean-field driver $H_0$ and the equation for the CC-\textit{bra}, are consistent.
The similarity is not surprising considering the fact that thermofield dynamics offers a unique way of computing thermal traces, much like the thermal Wick's theorem.

However, an explicit wave function representation of the thermal density matrix renders the thermofield approach far more versatile.
For instance, Ancilla DMRG, which is also based on the purification of an ensemble density matrix, is a thermal generalization of the matrix product states.
Similarly, wave functions such as multi-reference CI and CC and active space CI, for which diagrammatic equations and Wick's theorem are not straightforward, can also be extended to finite temperatures using our formalism. This is currently a work in progress.
Moreover, the different thermal CC methods discussed above work in the grand canonical ensemble. While the finite-temperature Wick's theorem and diagrammatic perturbation theory do not have an analog in the canonical ensemble, the wave function in thermofield dynamics is amenable to ground-state number projection techniques, such that a canonical ensemble formulation is straightforward and has been explored by the authors in Ref.~\citenum{harsha_wave_2020}.

\section{Conclusions}
In this short review, we have presented recent developments in finite-temperature wave function theories, focussing particularly on the formalism inspired by thermofield dynamics.
Within this approach, finite-temperature generalizations of the whole hierarchy of ground-state methods can be constructed.
We have presented benchmark calculations on a simple, yet non-trivial, one-dimensional Hubbard model. By comparing errors in internal energy, grand-potential and correlation functions, we confirm that finite-temperature wave functions behave similarly to their ground-state counterparts.
For moderately correlated electronic systems, both thermal CI and CC truncated at single and double excitations provide a significant improvement over mean-field.
While our implementation is capable of studying larger and more realistic electronic systems, the absence of exact benchmark data makes it difficult to assess the performance of the thermal wave function theories.
A comparative study involving different finite-temperature methods would be a natural step forward.

The explicit wave function representation of the finite-temperature Gibbs-Boltzmann density matrix allows the potential of extending this formalism not only to multi-reference methods, but also to emerging algorithms for quantum computers.
Application of, say, thermal CC to periodic systems is another venue that deserves more attention. It would largely facilitate the study of temperature-correlation phase diagrams in realistic materials.
Due to its size-extensive property, we expect thermal CCSD to perform well, particularly in comparison to mean-field and CISD, as we go to larger systems.
The finite-temperature methods inherit most of the features from their ground-state counterparts such as computational scaling, dependence of CC and CI on the reference state, etc.
Accordingly, we expect that thermal CC and CI theories would be most effective for weakly correlated systems.
The emerging field of finite-temperature wave function theories shows a lot of promise to compete with, as well as to supplement, presently standard methods such as quantum Monte Carlo and Matsubara perturbation theory.

\begin{acknowledgments}
  This work was supported by the U.S. Department of Energy, Office of Basic Energy Sciences, Computational and Theoretical Chemistry Program under Award No. DE-FG02-09ER16053. G.E.S. acknowledges support as a Welch Foundation Chair (No. C-0036).
\end{acknowledgments}

\bibliography{ThermalCC}

\end{document}